\documentclass[conference]{IEEEtran}
\usepackage[margin=0.75in]{geometry}
\usepackage{graphicx}
\usepackage{amssymb}
\usepackage{amsmath}
\usepackage{cases}
\usepackage{epstopdf}
\usepackage[english]{babel}
\usepackage[autostyle]{csquotes}
\title  {On the Capacity of Symmetric Gaussian 
Interference Channels with Feedback } 
\author{\IEEEauthorblockN{Lan V. Truong}
\IEEEauthorblockA{Information Technology Specialization Department (ITS)\\
FPT University, Hanoi, Vietnam\\
E-mail: lantv@fpt.edu.vn}
\and
\IEEEauthorblockN{Hirosuke Yamamoto}
\IEEEauthorblockA{Dept.~of Complexity Science and Engineering\\
The University of Tokyo, Japan\\
E-mail: hirosuke@ieee.org}
}
\date{}                                      
\begin{document}
\maketitle {}
\begin{abstract}                                                           
In this paper, we propose a new coding scheme for symmetric Gaussian interference channels with feedback based on the ideas of time-varying coding schemes. The proposed scheme improves the Suh-Tse and Kramer inner bounds of the channel capacity for the cases of weak and not very strong interference. This improvement is more significant when the signal-to-noise ratio (SNR) is not very high. It is shown theoretically and numerically that our coding scheme can outperform the Kramer code. In addition, the generalized degrees-of-freedom of our proposed coding scheme is equal to the Suh-Tse scheme in the strong interference case. The numerical results show that our coding scheme can attain better performance than the Suh-Tse coding scheme for all channel parameters. Furthermore, the simplicity of the encoding/decoding algorithms is another strong point of our proposed coding scheme compared with the Suh-Tse coding scheme. More importantly, our results show that an optimal coding scheme for the symmetric Gaussian interference channels with feedback can be achieved by using only marginal posterior distributions under a better cooperation strategy between transmitters.
\end{abstract}   
\begin{IEEEkeywords} Gaussian Interference Channel, Feedback, Posterior Matching, Iterated Function Systems.
\end{IEEEkeywords}           
\section{Introduction}
The capacity of the interference channel with feedback has been still unknown for many decades although there were some progresses toward solving this problem. Kramer developed a feedback strategy and derived an outer bound of the Gaussian channel [6], [9]. However, the gap between the outer and the inner bounds becomes large unboundedly as signal to noise ratio (SNR) and interference to noise ratio (INR) increase. Furthermore, Suh and Tse [1], [2] characterized the capacity region within 2 bits/s/Hz and the symmetric capacity within 1 bit/s/Hz for the two-user Gaussian interference channel with feedback. They also indicated that feedback provides \emph{multiplicative} gain at high SNR. However, their coding scheme does not work well when the SNR is close to the INR. Its symmetric coding rate becomes less than the Kramer code when this condition happens. In addition, it also has lower performance than the Kramer code when $\alpha = \log$ INR$/\log$ SNR is not very large and the SNR is low (cf. Figs. 1-2 of this paper or Fig. 14 in [2]). Later, the Suh-Tse coding scheme was extended to $M$-user Gaussian interference channels with feedback for $M \geq 3$ [7].
 \begin{figure}[ht]
	\centering
		\includegraphics[width=0.45\textwidth]{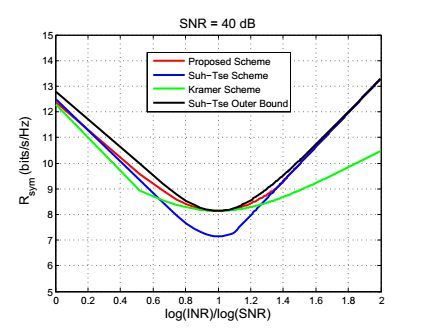}
	\caption{Symmetric Rate Comparision at High SNR}
	\label{fig:highsnr}
\end{figure}
\begin{figure}[ht]
	\centering
		\includegraphics[width=0.45\textwidth]{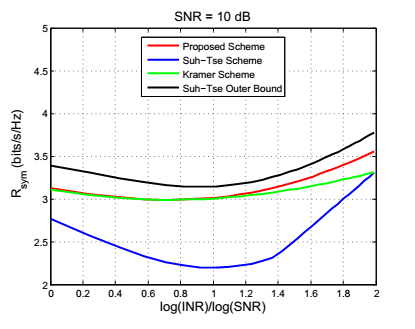}
	\caption{Symmetric Rate Comparision at Low SNR}
	\label{fig:lowsnr}
\end{figure}

In this paper, we propose a coding scheme which achieves better performance than the Suh-Tse code when $\alpha = \log$ INR$/\log$ SNR is not very large (See Figs.1-2 of this paper). In addition, our code can attain better symmetric rate than the Kramer code for all channel parameters, and therefore it overcomes all the weak-points of the Suh-Tse coding scheme and improves the Suh-Tse and Kramer inner bounds. For the strong interference case, our code can achieve the same generalized degrees-of-freedom as the Suh-Tse coding scheme. Furthermore, numerical results show that our coding scheme indeed has better/equal performance than/to the Suh-Tse coding scheme for all channel parameters.

\begin{figure}[ht]
	\centering
		\includegraphics[width=0.45\textwidth]{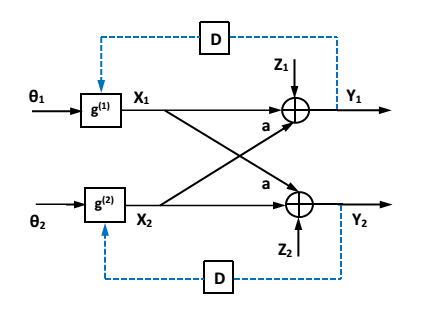}
	\caption{Gaussian Interference Channel with Feedback}
	\label{fig:icchannel}
\end{figure}
\section{Channel Model}
We consider the Gaussian interference channel shown in Fig. 3, which has two senders and two receivers. Sender 1 sends a source of information message points $\Theta_1$, which is uniformly distributed in (0,1), to receiver 1. Sender 2 sends a source of information points $\Theta_2$, which is also uniformly distributed in (0,1), to receiver 2. Assuming that $\Theta_1$ is independent of $\Theta_2$ and each channel interferes with the other. Specially, we assume that
\[
Y_1= X_1+ aX_2 + Z_1,
\]
\[
Y_2= X_2 + aX_1 + Z_2,
\]
where $Z_1 \sim \mathcal{N}(0,\sigma_1^2)$ and $Z_2 \sim \mathcal{N}(0, \sigma_2^2)$ are Gaussian noise random variables, and the input power constraints are $P_1$ and $P_2$, respectively. In this paper, we consider the symmetric interference channel such that $P_1 = P_2 =P$ and $\sigma_1^2 = \sigma_2^2 =1$. Hence, the signal-to-noise ratio and interference-to-noise ratio can capture channel gains SNR = $P$ and INR = $a^2P$. We also assume that output symbols are casually feedbacked to the corresponding sender and the transmitted symbol $X^{(m)}_n$ at time $n$ can depend on both the message $\Theta_m$ and the previous channel output sequence ${\bf Y}^{(n-1,m)}:=(Y_1^{(m)}, Y_2^{(m)},...,Y_{n-1}^{(m)})$ for $m \in \{1,2\}$. 

A \emph{transmission scheme} for the two-user Gaussian interference channel with feedback is sequences of measurable functions $\{g^{(m)}_n: (0,1) \times \mathbb{R}^{n-1} \rightarrow \mathbb{R}\}_{n=1}^{\infty}, m\in\{1,2\}$ so that the input to the channel generated by the transmitter is given by
\[
X^{(m)}_n = g^{(m)}_n(\Theta_m,{\bf Y}^{(n-1,m)}).
\]

A \emph{decoding rule} for the two-user Gaussian interference channel with feedback are sequences of measurable mappings $\{\Delta_n^{(m)}: \mathbb{R}^n \rightarrow \mathcal{E}\}_{n=1}^{\infty}, m\in \{1,2\}$ where $\mathcal{E}$ is the set of all open intervals in $(0,1)$ and $\Delta_n^{(m)}(y^{(n,m)})$ refers to the decoded interval at receiver $m$. The error probabilities at time $n$ associated with a transmission scheme and a decoding rule, is defined as
\[
p_n^{(m)}(e):=\mathbb{P}(\Theta_m \notin \Delta_n^{(m)}({\bf Y}^{(n,m)})), m\in \{1,2\},
\]
and the corresponding rate pair $(R_n^{(1)}, R_n^{(2)})$ at time $n$ is defined by
\[
 R_n^{(m)}:=-\frac{1}{n}\log\left|\Delta_n^{(m)}\left({\bf Y}^{(n,m)}\right)\right|.
\]
We say that a transmission scheme together with a decoding rule achieves a rate pair $(R_1, R_2)$ over a Gaussian interference channel if for $m\in \{1,2\}$ we have
\begin{equation}
\lim_{n\rightarrow \infty}\mathbb{P}\left(R^{(m)}_n<R_m\right)=0, \lim_{n \rightarrow \infty}p_n^{(m)}(e)=0.
\end{equation}
The rate pair is achieved within input power constraints $P_1, P_2$ if the following is satisfied:
\begin{equation}
\limsup_{n\rightarrow \infty}\frac{1}{n}\sum_{k=1}^n E[X^{(m)}_k]^2 \leq P_m, m \in \{1,2\}.
\end{equation}
The symmetric capacity is defined by $C_{sym}:= \sup\{R: (R,R) \hspace{1mm}\mbox{is achievable}\}$.

An \emph{optimal fixed rate} decoding rule for the two-user Gaussian interference channel with feedback for rate pair $(R_1,R_2)$ is the one that decodes a pair of fixed length intervals $\{(J_1,J_2): |J_m|=2^{-nR_m} \hspace{1mm}\mbox{for}\hspace{1mm} m\in\{1,2\}\}$, which maximizes posteriori probabilities, i.e.,
\[
\triangle^{(m)}_n(y^{(n,m)})=\underset{{J_m \in \mathcal{E}: |J_m|=2^{-nR_m}}}{\mbox{argmax}} \mathbb{P}_{\Theta_m|Y^n}(J_m|y^{(n,m)}).
\]
It is easy to see that the optimal fixed rate decoding rule for the Gaussian interference channel with feedback is the traditional MAP, MMSE decoding rule.

An \emph{optimal variable rate} decoding rule with target error probabilities $p^{(m)}_e(n)=\delta^{(m)}_n$ is the one that decodes a pair of minimal-length intervals $(J_1,J_2)$ such that accumulated marginal posteriori probabilities exceeds corresponding targets, i.e.,
\[
\triangle^{(m)}_n(y^{(n,m)})=\underset{{J_m\in \mathcal{E}: \mathbb{P}_{\Theta_m|Y^n}(J_m|y^{(n,m)})\geq 1- \delta^{(m)}_n}}{\mbox{min}}|J_m|.
\]
Both decoding rules use the marginal posterior distribution of the message point $\mathbb{P}_{\Theta_m|Y^n}$ which can be calculated online at the transmitters and the receivers.  Refer [6] for more details. \\

{\bf Lemma 1:} The achievability in the definition (1) and (2) implies the achievability in the standard framework.
\begin{IEEEproof}
See the detailed proof in [3], [4], and [5].
\end{IEEEproof}

\emph{Notations:} The cumulative distribution function (c.d.f.) of a random variable $X$ is given by $F_X(x)=\mathbb{P}_X((-\infty,x])$, and their inverse c.d.f. is defined as $F_X^{-1}(t):= \mbox{inf}\{x:F_X(x) > t\}$. The composition function is defined by $(f\circ g)(x)=f(g(x))$, and the sign function is defined as $\mbox{sgn}(x):=1$ if $x\geq 0$ and $\mbox{sgn}(x):=-1$ if $x<0$. We also use $(x)^+:=\max(x,0)$ and $\log^+(x):=\max(\log x,0)$.
\section{A Coding Scheme for Gaussian Interference Channels with Feedback}
In this section, we propose a \emph{time-varying} coding scheme for the symmetric Gaussian interference channel with feedback as following:
\subsection{Encoding}
\begin{itemize}
\item Step 1:
Transmitter $m$ sends $X^{(m)}_1=F_{X}^{-1}(\Theta_m)$ where $m\in \{1, 2\}$ and $X \sim \mathcal{N}(0,P_1)$ for some $P_1 >0$. We also set
\[
\rho_1:=\frac{E[F_{X}^{-1}(\Theta_1)F_{X}^{-1}(\Theta_2)]}{P_1} = 0.
\]
\item Step $n+1$ for $n \geq 1$:\\
Both transmitters estimate
\[
\rho_{n+1} = \frac{1}{\beta_n^2} \left\{ \rho_n - 2 b_n \mbox{sgn}(\rho_n) (|\rho_n| + |a|) \right. 
\]
\[
\left. +  b_n^2 \mbox{sgn}(\rho_n) [|\rho_n|(1+|a|^2) + 2 |a|]\right\}.
\]
Transmitter 1 sends $X_{n+1}^{(1)} \mbox{sgn}(\rho_{n+1})$ where
\[
X_{n+1}^{(1)}:=\frac{1}{\beta_n}(X_n^{(1)}- b_n \mbox{sgn}(\rho_n) Y_n^{(1)}).
\]
Transmitter 2 sends $X_{n+1}^{(2)}\mbox{sgn}(a)$ where
\[
X_{n+1}^{(2)}:=\frac{1}{\beta_n}(X_n^{(2)}- b_n \mbox{sgn}(a) Y_n^{(2)}).
\]
Receiver 1 receives
\[
Y_{n+1}^{(1)} = X_{n+1}^{(1)}\mbox{sgn}(\rho_{n+1}) + |a| X_{n+1}^{(2)} + Z_{n+1}^{(1)}.
\]
Receiver 2 receives
\[
Y_{n+1}^{(2)} =\mbox{sgn}(a) X_{n+1}^{(2)} + a \hspace{1mm} \mbox{sgn}(\rho_{n+1}) X_{n+1}^{(1)} + Z_{n+1}^{(2)}.
\]
Both receivers feedback their received signals to the corresponding transmitters.\\
Here, $(\beta_n > 0, b_n)$ should be chosen to satisfy the following constraints:
\begin{equation}
\limsup_{N\rightarrow \infty} \frac{1}{N} \sum_{n=1}^N E[X_n^{(m)}]^2 \leq P, \hspace{1mm}\forall m\in\{1,2\}.
\end{equation}
\end{itemize}
\subsection{Decoding}
 \begin{itemize}
\item At each time slot $n$, receiver $m \in \{1,2\}$ selects a fixed interval $J_1^{(m)}=(s_m,t_m) \subset \mathbb{R}$ as the decoded interval with respect to $X_{n+1}^{(m)}$.
\item Then, set the decoded interval  $J_n^{(m)}=\left(T^{(m)}_n(s_m),T^{(m)}_n(t_m)\right)$,  as the decoded interval with respect to $X_1^{(m)}$, where  
 \[
 T^{(m)}_n(s):= w^{(m)}_1\circ w^{(m)}_2\circ \cdots \circ w^{(m)}_n (s), \hspace{2mm} \forall s\in \mathbb{R}
 \]
and 
\[
w^{(1)}_n:=\beta_n x + b_n \mbox{sgn}(\rho_n) Y_n^{(1)},
\]
\[
w^{(2)}_n:=\beta_n x + b_n \mbox{sgn}(a) Y_n^{(2)}.
\]
\item Receiver $m$ sets the decoded interval for the message $\Theta_m$ as follows:
\[
\Delta_n^{(m)}\left({\bf Y}^{(n,m)}\right)=F_{X_m}(J_n^{(m)}).
\] 
\end{itemize}

We call this coding strategy the \emph{Gaussian interference time-varying feedback coding strategy}, which is an \emph{optimal variable rate} decoding rule with doubly exponential decay of targeted error probabilities  (see the proof of the Lemma 2 in this paper). 
\section{A New Achievable Rate Region}
{\bf Lemma 2:} Under the condition that $0< \limsup_{n\rightarrow \infty} \beta_n < 1$, the \emph{time-varying coding scheme}  for the symmetric Gaussian interference channel with feedback achieves the following symmetric rate:
\[
R_{sym} = -\limsup_{n\rightarrow \infty} \log \beta_n  \hspace{3mm}  \mbox{(bits/channel\hspace{1mm}use)}.
\]
\begin{IEEEproof} We provide a sketch of the proof. The detailed one can be found in papers [3], [4], and [5].
\begin{itemize}
\item Define $\beta:=\limsup_{n\rightarrow \infty} \beta_n$. It is easy to see that $R_{sym} = \log \beta^{-1}$. 
\item For any $R  < R_{sym} $, we can find an $\epsilon >0$ such that $R  < \log(\beta +\epsilon)^{-1}$.
\item Choose an $N_{\epsilon} \in \mathbb{N}$ such that $\sup_{n\geq N_{\epsilon}}\beta_n < \beta + \epsilon$.
\item Using the Law of Iterated Expectations (Fubini's Theorem), we can show that
\[
\mathbb{P}(R^{(m)}_n < R) \leq A_{\epsilon} 2^{nR}(\beta+\epsilon)^{(n-N_{\epsilon})}|J_1^{(m)}|.
\]
\item We can also show that
\[
p_n^{(m)}(e)=O\left(\exp\left(-\frac{|J_1^{(m)}|^2}{8P}\right)\right).
\]
By choosing $|J_1^{(m)}|=o\left(2^{n(\log(\beta+\epsilon)^{-1} -R)}\right) \approx o\left(2^{n(R_{sym} -R)} \right)$, and $t_m=-s_m = |J_1^{(m)}|/2$, the aforementioned condition (1) is satisfied. Here, symbols $O(x)$ and $o(x)$ are Landau symbols. Besides, the choice of sequences $(b_n, \beta_n)$ following the rule (3) leads to the fact that the input power constraints are also satisfied. 
\end{itemize}
\end{IEEEproof}
{\bf Theorem 1:}
The non-degraded symmetric Gaussian interference channel ($a \neq 0$) can achieve the following symmetric rate:
\[
R_{sym} \mbox{(bits/channel\hspace{1mm}use)}=\frac{1}{2}\max_{\rho \in [0,\rho_0], \hspace{1mm} b \in \{b_1^*, b_2^*\}}\times
\]
\[
\times  \log\left[\frac{P}{P+b^2[1+P + |a|^2 P + 2|a| P \rho] - 2 bP[1+|a|\rho]}  \right],
\]
where 
\[
0< \rho_0: =\sqrt{\frac{a^2P^2 + P -\sqrt{P[2a^2P^2 +P]}}{a^2P^2}}<1,
\]
and
\[
b^*_{1,2} = \frac{2P\rho + |a|P + |a|P\rho^2}{2|a|P + 2P\rho + 2|a|^2 P \rho + \rho + 2|a|P \rho^2} 
\]
\[
\pm \frac{\sqrt{P^2 |a|^2 \rho^4 - 2\rho^2(|a|^2P^2 +P)+|a|^2 P^2}}{2|a|P + 2P\rho + 2|a|^2 P \rho + \rho + 2|a|P \rho^2}.
\]
\begin{IEEEproof}
From our transmission strategy, we have
\begin{equation}
X_{n+1}^{(1)} =\frac{1}{\beta_n}(X_n^{(1)}- b_n \mbox{sgn}(\rho_n) Y_n^{(1)}),
\end{equation}
\begin{equation}
X_{n+1}^{(2)} =\frac{1}{\beta_n}(X_n^{(2)}- b_n \mbox{sgn}(a) Y_n^{(2)}).
\end{equation} Denote
\[
P_n=E[X^{(1)}_n]^2 = E[X^{(2)}_n]^2, \rho_n = \frac{E[X_n^{(1)} X_n^{(2)}]}{P_n}.
\]
(We can show by induction that $E[X^{(1)}_n]^2 = E[X^{(2)}_n]^2$ for all $n$). Therefore, it is easy to see that
\[
E[X_n^{(2)} Y_n^{(1)}] = P_n(|\rho_n| + |a|),
\]
\[
E[X_n^{(1)} Y_n^{(2)}] =P_n\mbox{sgn}(a) \mbox{sgn}(\rho_n) (|\rho_n| + |a|),
\]
\[
E[Y_n^{(1)} Y_n^{(2)}] = P_n \mbox{sgn}(a) \left[|\rho_n| (1+ |a|^2) + 2|a|\right].
\]
Note that
\[
E[X_{n+1}^{(1)} X_{n+1}^{(2)}] = \frac{1}{\beta_n^2} \left\{E[X_n^{(1)} X_n^{(2)}]\right.
\]
\[
 \left. - b_n\mbox{sgn}(\rho_n) E[X_n^{(2)} Y_n^{(1)}] - b_n \mbox{sgn}(a) E[X_n^{(1)} Y_n^{(2)}]\right.
\]
\[
\left. + b_n^2 \mbox{sgn}(\rho_n)\mbox{sgn}(a) E[Y_n^{(1)} Y_n^{(2)}]\right\}.
\]
Finally, we obtain
\[
P_{n+1}\rho_{n+1} =P_n \mbox{sgn}(\rho_n) \frac{1}{\beta_n^2} \left\{|\rho_n| - 2 b_n  (|\rho_n| + |a|) \right.
\]
\begin{equation}
\left. + b_n^2  [|\rho_n|(1+|a|^2) + 2 |a|]\right\}.
\end{equation}
Similarly, observe that
\[
E[X_n^{(1)} Y_n^{(1)}] =  P_n \mbox{sgn}(\rho_n) [1 + |a| |\rho_n|],
\]
\[
E[X_n^{(2)} Y_n^{(2)}] =  P_n \mbox{sgn}(a)[1 + |a| |\rho_n|],
\]
\[
E[Y_n^{(1)}]^2 =  1 + P_n + a^2 P_n + 2|a| |\rho_n| P_n,
\]
\[
E[Y_n^{(2)}]^2 = 1 + P_n + a^2 P_n + 2 |a| |\rho_n| P_n.
\]
From the relations (4) and (5) we have
\[
P_{n+1} = \frac{1}{\beta_n^2} \left\{ P_n - 2 P_n b_n [1 + |a| |\rho_n|] \right.
\]
\begin{equation}
\left. + b_n^2 \left[1 + P_n + a^2 P_n + 2|a| |\rho_n| P_n\right] \right\}.
\end{equation}
From (6) and (7), if we can force $P_n\rightarrow P, b_n \rightarrow b, \beta_n \rightarrow \beta$ and $|\rho_n| \rightarrow \rho \in [0,1]$, we obtain the following equations with three unknowns $(b, \rho, \beta)$:
\small
\begin{subequations} \label{group4}
\begin{align}
P &= \frac{1}{\beta^2} \left[ P - 2 b P (1 + |a| \rho) + b^2 (1 + P + a^2 P + 2 |a| \rho P)\right] \label{x4}, \\
-\rho &= \frac{1}{\beta^2} \left[\rho - 2b(\rho +|a|) + b^2\left(\rho(1+|a|^2) + 2 |a|\right)  \right]. \label{y4}
\end{align}
\end{subequations}
\normalsize
By eliminating $\beta$ and considering $\rho$ as a running variable, we have the following quadratic equation in $b$ for each fixed choice of $\rho$:
\[
b^2[2|a| P + 2P\rho + 2|a|^2 P \rho + \rho + 2|a|P \rho^2] 
\]
\begin{equation}
- 2b[2P\rho + |a| P + P|a|\rho^2] + 2P\rho =0.
\end{equation}
After some simple derivations, the discriminant of this quadratic equation is given by
\begin{equation}
\Delta = P^2 |a|^2 \rho^4 - 2\rho^2(|a|^2 P^2 +P) +|a|^2 P^2: = f(\rho).
\end{equation}
Since $f(0) = |a|^2 P^2 > 0 $ and $f(1) = -2P <0$, there exists the minimum value $\rho_0 \in (0,1)$ such that $f(\rho_0)=0$. Specifically, the value of $\rho_0$ is given by
\[
0< \rho_0 =\sqrt{\frac{|a|^2P^2 + P -\sqrt{P[2a^2P^2 +P]}}{a^2P^2}}<1.
\]
On the other hand, since the derivative of $f(\rho)$ satisfies
\[
f'(\rho) =  4P^2 |a|^2 (\rho^3-\rho) - 4P\rho \leq 0,
\] for all $\rho \in [0,1)$, we have $\Delta = f(\rho) \geq 0$  for all $\rho \in [0, \rho_0]$. For all these values of $\rho$, we can easily show that (9) can have two positive solutions $b^*_1, b^*_2$ as the theorem statement. As a result, (8a) and (8b) have at least one solution $(b,\beta)$ for each fixed  $\rho \in [0,\rho_0]$. Therefore,  if we force all the sequences $(b_n, \beta_n, P_n, |\rho_n|)$ to converges to $(b, \beta, P, \rho)$, the symmetric Gaussian interference channel with feedback can achieve the following rate by the Lemma 2:
\[
R_{sym}= \left(- \log(\min_{\rho \in [0,\rho_0]} \beta)\right)^+ = \frac{1}{2} \max_{\rho \in [0,\rho_0], b \in \{b_1^*, b_2^*\} }\times 
\]
\[
\log^+\left(\frac{P}{P+b^2[1+P + |a|^2 P + 2|a| P \rho] - 2 bP[1+|a|\rho]}  \right)
\]
Note that since Lemma 2 holds only for $0< \beta <1$, the superscript $+$ is necessary to deal with general cases.

To complete the proof, we show a procedure to force $|\rho_n| = \rho \hspace{1mm}(\rho_n = (-1)^n \rho)$ for any $\rho \in [0, \rho_0]$, $b_n = b$, $P_n =P$, $\beta_n= \beta$ for any $n\geq 2$.
Indeed, from (6) and (7), we firstly force $P_2=P, \rho_2 = \rho$, and $\beta_2=\beta$ by setting
\begin{equation}
P\rho=\frac{P_1}{\beta_1^2}\{ 2|a|(b_1^2-b_1)\},
\end{equation}
\begin{equation}
P= \frac{1}{\beta_1^2} \{P_1- 2b_1P_1 + b_1^2(1+P_1+a^2P_1)\}.
\end{equation}
This procedure is feasible because (11) and (12) have at least one solution which is a triplet $(b_1, P_1 >0, \beta_1>0)$ for each $\rho \in [0,\rho_0]$. Indeed,
\begin{itemize}
\item For $\rho=0$, we can choose $b_1=0, P_1 =P, \beta_1=1$.
\item For $\rho \neq 0$, from (11) and (12) we have the following quadratic equation in $b_1$
\end{itemize}
\begin{equation}
b_1^2[(1+P_1+a^2P_1)\rho-2|a|P_1] - 2(\rho-|a|)P_1 b_1 + P_1 \rho=0.
\end{equation}
The discriminant of this quadratic equation can easily be shown to be
equal to $E(\rho)= a^2(1-\rho^2)P_1^2 - P_1 \rho^2$. For the case $\rho \neq |a|$, we can choose $P_1$ such that this discriminant is equal to zero by setting $P_1 =\rho^2/(a^2(1-\rho^2))$. By this choice of $P_1$, we obtain
\[
b_1 = \frac{(\rho-|a|)P_1}{(1+P_1+a^2P_1)\rho-2|a|P_1}.
\] 
In order for (11) and (12) to have solution $\beta_1$, we need to show that the above choices of $P_1, b_1$ satisfy $b_1^2 - b_1 >0$. Clearly for $\rho< |a|$, this requirement is satisfied by noting that $b_1<0$ since $\rho-|a|<0$ and 
\[
\rho > \frac{2|a|\rho^2}{a^2+\rho^2} = \frac{2|a|P_1}{1+P_1+a^2P_1}.
\]
For $\rho > |a|$ ($|a|<1$, of course), observe that
\[
P_1 = \frac{\rho^2}{a^2(1-\rho^2)} > \frac{\rho}{|a|-a^2 \rho}.
\]
It follows that
$
(\rho-|a|)P_1 > (1+P_1+a^2 P_1)\rho - 2|a|P_1 >0
$
or $b_1 >1$. This also means that $b_1^2 - b_1 >0$. For the case $\rho = |a|$ any choice of $P_1 > 1/(1-a^2)$ works since we have $E(|a|)>0$. Besides, the sum of two solutions of the quadratic equation (13) in $b_1$ is equal to zero, and their product is not equal to zero (by using Vieta's formula). Hence, we must find at least one $b_1<0$ or $b_1^2 - b_1 >0$.

Finally, we only need to set $\beta_n =\beta, P_n =P, b_n = b$ which is a solution of (8a) and (8b) for each choice of $\rho \in [0,\rho_0]$ and for all $n\geq 3$. Here $b$ should be chosen to minimize $\beta$ for each choice of $\rho \in [0,\rho_0]$ in order to maximize the achievable symmetric rate of our coding scheme. Last but not least, we can show that our code has better performance than the Kramer code [6], and therefore the superscript $+$ can be got rid of from the achievable symmetric rate formula.
\end{IEEEproof}
\emph{Remark 1:} For the degraded Gaussian interference channel with feedback $(a=0)$, (8a) and (8b) become
\begin{equation}
-\rho = \frac{1}{\beta^2} \rho (b-1)^2, P =\frac{1}{\beta^2}[P(b-1)^2 + b^2].
\end{equation}
From (14), we must have $\rho =0$ and the achievable symmetric rate becomes
\[
R_{sym} =\frac{1}{2}\max_{b \in \mathbb{R}} \log \left[\frac{P}{P(b-1)^2 + b^2}\right]= \frac{1}{2}\log(1+P).
\]
This result coincides with the well-known capacity of this channel with no interference.

{\bf Corollary 1:} The proposed \emph{time-varying code} outperforms the Kramer code for all channel parameters. 
\begin{IEEEproof}
A variant of Kramer code is constructed by choosing triplet $(b,\beta, \rho)$, which is a solution of (8a) and (8b), as follows:
\[
b= \frac{P(1+|a|\rho)}{P(1+a^2+2|a|\rho)+1},
\beta = \sqrt{\frac{a^2P(1-\rho^2)+1}{P(1+a^2+2|a|\rho)+1}},
\]
and $\rho$ is the unique solution in $(0,1)$ of the next equation:
\[
2|a|^3P^2\rho^4 + a^2P\rho^3-4|a|P(a^2P+1)\rho^2
\]
\begin{equation}
-(2a^2P+P+2)\rho + 2|a|P(a^2P+1)=0.
\end{equation} 
(See also in [2], [6], [9]). Therefore, it is inferior to the proposed code in this paper.

\emph{Remark 2:} The choice of $b$ and $\beta$ for the Kramer code is to maximize the achievable rate (or minimize $\beta$) by using only (8a)  for each fixed value of $\rho$. In order to satisfy (8b), the choice of $b$, $\beta$ in the above proof is applicable only to the fixed value of $\rho$ which is the unique solution in $(0,1)$ of (14). For other values of $\rho$, from (8a) and (8b), we see that $b$ and $\beta$ are given by two different functions of $\rho$. Enlarging the set of possible choices of $\rho$ (then $b$, $\beta$) increases the achievable symmetric rate in this paper.
\end{IEEEproof}

{\bf Corrolary 2:} For $\alpha = \log INR/ \log SNR > 1$, the generalized degrees of freedom of the proposed coding scheme is given by
\[
d(\alpha):=\lim_{SNR, INR \rightarrow \infty} \frac{R_{sym}(SNR, INR)}{\log SNR}= \frac{\alpha}{2}.
\]
(Here, the unit of $R_{sym}$ is bits/s/Hz). As a consequence, the proposed coding scheme has the same generalized degrees of freedom as the 
Suh-Tse coding scheme [1], [2] and also provides the \emph{multiplicative} gain at high SNR. Note that the Kramer code achieves only $d(\alpha) = (1+\alpha)/4$ for $\alpha \geq 1$ (cf. (48) [2]).
\begin{IEEEproof}
Since $\alpha = \log INR/ \log SNR > 1$, we have $|a|^2 P = P^{\alpha}$, or $|a| = P^{(\alpha-1)/2}$. For $P$ sufficiently large and $\rho \approx 1$, observe that
\[
b^*_{1,2}\approx \frac{P^{(\alpha+1)/2}(1+\rho^2)}{2P^{\alpha}\rho} \pm \frac{P^{(\alpha+1)/2}(1-\rho^2)}{2P^{\alpha}\rho}. 
\]
By choosing $b=b^*_1 \approx  P^{(1-\alpha)/2}\rho$, we obtain
\[
R_{sym}(\rho)  \approx \log^+ \left( \frac{P}{P+b^2 P^{\alpha} - 2bP(1+P^{(\alpha-1)/2}\rho)}\right)
\]
\begin{equation}
=-\log\left(1-\rho^2 -2\rho P^{(1-\alpha)/2}\right) \hspace{2mm} \mbox{(bits/s/Hz)}.
\end{equation}
From Theorem 1, we also have
\[
\rho_0^2 =1+P^{-\alpha}-\sqrt{2 P^{-\alpha}+  P^{- 2 \alpha}} > 1- \sqrt{3} P^{-\alpha/2},
\]
and the $\rho_*$, that maximizes the achievable rate, must be in the interval $[0,\rho_0]$. This condition is satisfied by setting $\rho_* =  \sqrt{1-P^{-\gamma}+P^{-(\alpha-1)}} - P^{-(\alpha-1)/2}$ for an arbitrary positive number $\gamma <\alpha/2$ although this $\rho_*$ may not be optimal. Indeed,  for $P$ sufficiently large, we have $\rho_* \approx 1$ and 
\[
\rho_* < \sqrt{1-P^{-\gamma}}+ \sqrt{P^{-(\alpha-1)}} - P^{-(\alpha-1)/2} =\sqrt{1-P^{-\gamma}}.
\] 
Hence $\rho_*^2 < 1- P^{-\gamma} < 1- \sqrt{3} P^{-\alpha/2} < \rho_0^2$. On the other hand,  since $1-\rho_*^2 -2\rho_* P^{(1-\alpha)/2} =  P^{-\gamma}$, we obtain
\[
d(\alpha)\geq \lim_{SNR, INR \rightarrow \infty}\frac{R_{sym}(\rho_*)}{\log SNR} = \gamma.
\]
Since $\gamma$ can take any arbitrary value less than $\alpha/2$, we have $d(\alpha) \geq \alpha/2$. From the result of [2], it is known that $d(\alpha) \leq \alpha/2$. Hence, we must have $d(\alpha)=\alpha/2$.  
\end{IEEEproof}

\section{Numerical Evaluation}
In order to evaluate $R_{sym}$ in Theorem 1 numerically, we determined the optimal $\rho$ by increasing it from zero by incremental step $10^{-5}$. The numerical results are shown in Figs. 1-2 in Section I. We note that our proposed code can achieve better/equal symmetric rate than/to the Suh-Tse and Kramer codes for all channel parameter $(a,P)$. This improvement is more significant when SNR is not too high. 
\section{Conclusion}
The inner bound of the capacity is improved compared with the Suh-Tse and Kramer inner bounds by analyzing the performance of our proposed code as an optimized form of the Kramer code. Our result also shows that an optimal coding scheme for the Gaussian interference channel with feedback can be achieved by using marginal posterior distributions.
\section*{Acknowledgment}
This work was supported in part by JSPS KAKENHI Grant Number 25289111. 


\begin{thebibliography}{1}
	\bibitem{key1}
Changho Suh and David Tse,
\enquote{Symmetric Feedback Capacity of the Gaussian Interference Channel to Within One Bit,}
in {\em Proc. Int. Symp. Information Theory}, Jun. 2009.
\bibitem{key2}
Changho Suh and David Tse,
\enquote{Feedback Capacity of the Gaussian Interference Channel to Within 2 Bits,}
{\em IEEE Trans. Inf. Theory}, vol. 57, No. 5, pp. 2667-2685, May 2011.
\bibitem{key3}
Lan V. Truong, \enquote{Posterior Matching Scheme for Gaussian Multiple Access Channel with Feedback,} [Online]. Available: http://arxiv.org/abs/1204.4249.
\bibitem{key4}
Lan V. Truong, \enquote{Posterior Matching Scheme for Gaussian Multiple Access Channel with Feedback,}
in {\em Proc. IEEE Information Theory Workshop}, Nov. 2014.
\bibitem{key5}
Lan V. Truong, \enquote{A Novel Time-Varying Coding Scheme for the Gaussian Broadcast Channel with Feedback,}.  [Online]. Available: http://arxiv.org/abs/1404.2520.
\bibitem{key6}
Gerhard Kramer,
\enquote{Feedback Strategies for White Gaussian Interference Networks,}
{\em IEEE Trans. Inf. Theory}, vol. 48, pp.1423-1438, Jan. 2002.
\bibitem{key7}
 Ravi Tandon, Soheil Mohajer, and H. Vincent Poor,
\enquote{On the Symmetric Feedback Capacity of the K-User Cyclic Z- Interference Channel,}
 {\em IEEE Trans. Inf. Theory}, vol. 59, no.5, pp. 2713-2733, May  2013.
\bibitem{key8}
 Ravi Tandon, Soheil Mohajer, and H. Vincent Poor,
\enquote{On the Feedback Capacity of the Fully Connected K-User Interference Channel,}
 {\em IEEE Trans. Inf. Theory}, vol. 59, no.5, pp. 2863-2881, May  2013.
\bibitem{key9}
G. Kramer, \enquote{Correction to \enquote{Feedback Strategies for White Gaussian
Interference Networks}, and a Capacity Theorem for
Gaussian Interference Channels with Feedback,} {\em IEEE Trans. Inf. Theory}, vol.
50, no. 6, pp. 1373-1374, Jun 2004.
\bibitem{key10}
O. Shayevitz and M. Feder, \enquote{Optimal Feedback Communication via Posterior Matching,}
\emph{IEEE Trans. Inf. Theory}, vol. 57, no.3, pp.1186-1221, Mar. 2011.
\end{thebibliography}
\end{document}